\documentclass[prx,aps,floatfix,twocolumn,preprintnumbers,superscriptaddress]{revtex4-1}
\usepackage{amsmath}
\usepackage{amsfonts}
\usepackage{amssymb}
\usepackage{epsfig}
\usepackage{graphicx}
\usepackage{color}

\begin{document}

\title{Destabilization of a flow focused suspension of magnetotactic bacteria}

\author{Nicolas Waisbord}
\affiliation{Institut Lumi\`ere Mati\`ere, CNRS UMR 5306, 
Universit\'e Claude Bernard Lyon1, Universit\'e de Lyon, France}

\author{ Christopher Lef\`evre}
\affiliation{CEA Cadarache/CNRS/Aix-Marseille Universit\'e, UMR7265 Institut de Biologie Environnementale et Biotechnologie, 13108, Saint Paul lez Durance, France}

\author{Lyd\'eric Bocquet}
\affiliation{Institut Lumi\`ere Mati\`ere, CNRS UMR 5306, 
Universit\'e Claude Bernard Lyon1, Universit\'e de Lyon, France} 

\author{Christophe Ybert}
\affiliation{Institut Lumi\`ere Mati\`ere, CNRS UMR 5306, 
Universit\'e Claude Bernard Lyon1, Universit\'e de Lyon, France} 

\author{C\' ecile Cottin-Bizonne}
\affiliation{Institut Lumi\`ere Mati\`ere, CNRS UMR 5306, 
Universit\'e Claude Bernard Lyon1, Universit\'e de Lyon, France}
\email{Cecile.Cottin-Bizonne@univ-lyon1.fr}

\begin{abstract}
{Active matter is a new class of material, intrinsically out-of equilibrium with intriguing properties. So far, the recent upsurge of studies has mostly focused on the spontaneous behavior of these systems --in the absence of external constraints or driving--. Yet, many real life systems evolve under constraints, being both submitted to flow and various taxis. In the present work, we demonstrate a new experimental system which opens up the way for quantitative investigations, and discriminating examinations, of the challenging theoretical description of such systems. We explore the behavior of magnetotactic bacteria as a particularly rich and versatile class of driven matter, which behavior can be studied under contrasting and contradicting stimuli. In particular we demonstrate that the competing driving of an orienting magnetic field and hydrodynamic flow lead not only to jetting, but also unveils a new pearling instability. This illustrates new structuring capabilities of driven active matter.}

 \end{abstract}

\maketitle

Active matter has been an emerging and very flourishing field in the past decade. This led to the development of many numerical and theoretical investigations, complemented by --much fewer-- experimental characterizations \cite{Cates2012}. So far, studies on active particles have mostly focused on the spontaneous behavior of individual particles or of assemblies of active particles with the emergence of clusters or phase separation. One can even mention that the non biased motion of the particles has been sometimes part of the definition of active matter \cite{Ramaswamy2010}.  Although bearing much significance, \textit{e.g.} from a fundamental statistical physics point of view, many relevant practical situations actually involve additional external constraints. This encompasses the ubiquitous presence of surrounding liquid flows with essential consequences spanning from biological systems like sperm cells \cite{Riffell2007} to the design of artificial micro-robots that would make true R. Feynman old dream of minimally invasive medicine \cite{Feynman1959}. In this situation, swinging and tumbling trajectories of spherical swimmers in Poiseuille flow have, for instance, been identified theoretically \cite{Zotl2012}. In addition, the possibility of a bacteria localization mechanisms in flows, due to the so-called shape-induced rheotaxis, has been demonstrated recently \cite{Marcos2012, Rusconi2014}.

Beyond their interaction with the surrounding hydrodynamic environment, a whole set of microswimmers possess the additional ability to bias their motion in response to an external stimuli (either passively --gravity, magnetic fields--, or actively --chemical, optical, light fields--). The ubiquitous ability of living active matter to bias its motion is of key importance in observed collective motions 
\cite{Pedley1992,Koch2011,Ezhilan2013}. The understanding and modeling of active systems behaviors thus critically requires a model experimental system of \textit{driven} active matter. This is the purpose of the present work, which introduces magnetotactic bacteria as a promising benchmark system of driven active matter that offers unprecedented capabilities for physical and quantitative investigations. Such class of bacteria synthesizes nanomagnets in their membrane, providing a permanent magnet of the order of  $M= 1 \times 10^{-16}{\rm A.m}^2$ \cite{Bazylinski2004, Meldrum1993}. These magnetosomes enable them to be very easily remotely controlled by a magnetic field, which orients the motion of the bacteria by a magnetic torque \cite{Martel2009}. The motion of driven magnetotactic bacteria,   \emph{Magnetococcus Marinus type 1} strain MC-1 \cite{Bazylinski2013}  is experimentally studied here when facing a steady Poiseuille flow. At small flow velocities this system displays a focusing into a magnetotactic jet, reminiscent of observation in gravitactic or phototactic systems \cite{Kessler85, Garcia13}, gathering both of the advantages of these two taxis: the physical simplicity of the action on the orientation of the first one and the easily and continuously tunable aspect of the second. Remarkably, this novel experimental bio-system is shown to quantitatively obey simple physical description of Fokker-Planck type. To our knowledge, this is the first quantitative testing of this wide-spread approach of active driven matter under flow. Increasing the flow velocities, we moreover report a new instability leading to the development of recirculating swarms of magnetotactic bacteria. This sheds some new light on the structuring capabilities of driven active matter with possible outcomes in the many promising microswimmers able to orient with homogeneous magnetic field \cite{Ghosh2009,Gao2014}.

\subsection*{I. ORIENTATION IN A MAGNETIC FIELD}
\begin{figure}
\centering
    \includegraphics[width=0.5\textwidth]{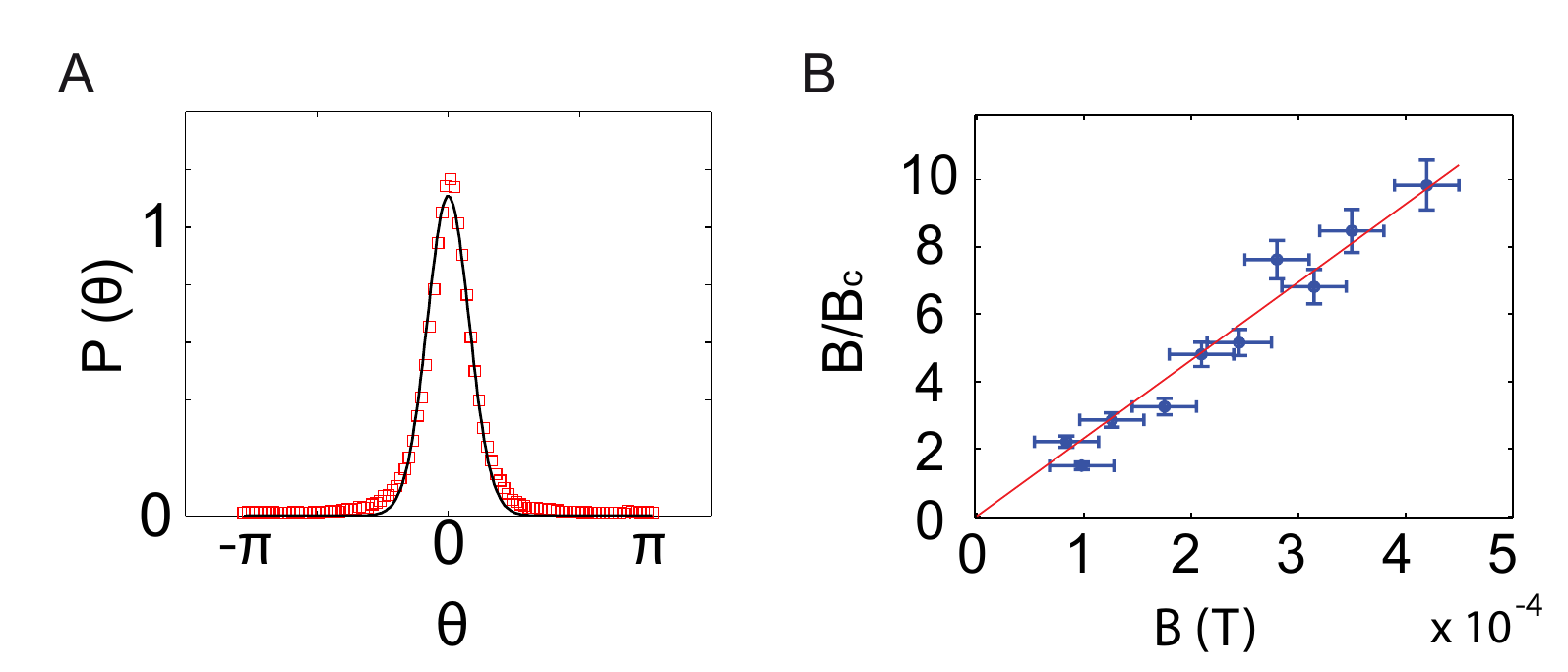}
    
    \caption{Bacteria orientation. A: Experimental statistics of orientations for a magnetic field of 0.42 mT adjusted with $P(\theta) \propto  e^{\alpha\cos(\theta)}$ as proposed by Langevin model (see text for details). B: Evolution of $\alpha$ with the magnetic field; the red line is the linear adjustment $\alpha=B/B_c$ with a slope $1/B_c = 2.3\pm 0.1\times 10^4  {\rm T^{-1}}$.\label{fig:fig0}}  
\end{figure}

As a mandatory step towards quantitative description of the external driving, we first characterize the ability of MC-1  bacteria to be directed by a magnetic field. MC-1 bacteria  are placed in a microfluidic chamber in a still standing fluid  (see Supplementary Materials), where their response to a \textit{uniform} magnetic field $\mathbf{B}$ is measured.
Qualitatively, randomly oriented bacteria with individual swimming velocity of \textbf{$V_\mathrm{swim}= 100 \pm 10 \mu {\rm m/s}$}, get oriented by  $\mathbf{B}$, and swim on average along the magnetic field lines. More quantitatively, for a given magnetic field, the statistics of orientations of bacteria (fig.\ref{fig:fig0}.A) show a distribution centered along the field direction, with a  width associated to a finite orientation noise. This experiment has long been considered as a way to estimate the magnetic momentum of bacteria assuming that only the thermal bath was at stake for the disorientation \cite{Blakemore1982, Nadkarni2013}. However a recent study suggests that other sources can be involved in the disorientation phenomena \cite{Wu2014}.

The actual random torque $\zeta(t)$ thus encompasses not only thermal fluctuations, but also the possibility of tumbling or swimming noise. Nevertheless, we hereafter assume that it all can be summed up in a total contribution which remains in structure uncorrelated in time \cite{Rusconi2014}: 
\begin{equation}
<\zeta(t)\zeta(t')> = \xi^2 \delta(t-t')  \rm{\ and} <\zeta(t)> = 0.
\end{equation}

This problem remains analogous to the classical Langevin treatment with the orientation of the bacteria set by the equilibrium between (i) the magnetic torque $-M B \sin\theta$
(with $\mathbf{M}$ the magnetic momentum of the bacterium  and $\theta$ the angle between the bacterium axis and $\mathbf{B}$), (ii)  the viscous torque $-{\dot{\theta}}/{\mu_R}$ (with $\mu_R = ({8\pi\eta r^3})^{-1}$ the mobility, $r$ the hydrodynamic radius and $\eta$ the viscosity of the liquid) and (iii) the random torque $\zeta(t)$.
 This predicts a probability orientation matching Langevin paramagnetism, with an \textit{a priori} unknown noise amplitude (see Supp. Mat.):
\begin{equation}
 P(\theta) \propto  e^{\frac{2MB}{\xi^2\mu_R}\times\cos \theta} \propto e^{\frac{B}{B_c} cos \theta},
 \label{eqP}
\end{equation}
with $B_c$ a characteristic magnetic field scale defined by
\begin{equation}
B_c = \frac{\xi^2\mu_R}{2M}.
\end{equation}
%

%
%%%%%%%%%%%%%%%%%%%%%%%%%%%%%%%
 %
 %%%%%%%%%%%%%%%%%%%%%%%%%%%%%%
 
As can be seen in fig.\ref{fig:fig0}.A, the orientation distribution of bacteria is very well-described by the Langevin prediction $\exp(\alpha\cos\theta)$ eq. (\ref{eqP}). Plotting $\alpha$ versus $B$ in fig.\ref{fig:fig0}.B we demonstrate the validity of the linear dependency in $B$ expected from eq. (\ref{eqP}) with $\alpha\equiv B/B_c$, and obtain $B_c = 43\pm 2 \mu {\rm T}$. Despite the rotational diffusion sources might be multiple in these living microswimmers, the balance with magnetic torque is hence fully summed up in this characteristic magnetic field that we can measure for magnetotactic bacteria, and that we might be able to choose for artificial magnetic microswimmers. 

%
%%%%%%%%%%%%%%%%%%%
\subsection*{II. FLOW FOCUSED SUSPENSION}

\begin{figure}
\centering

    \includegraphics[width=0.5\textwidth]{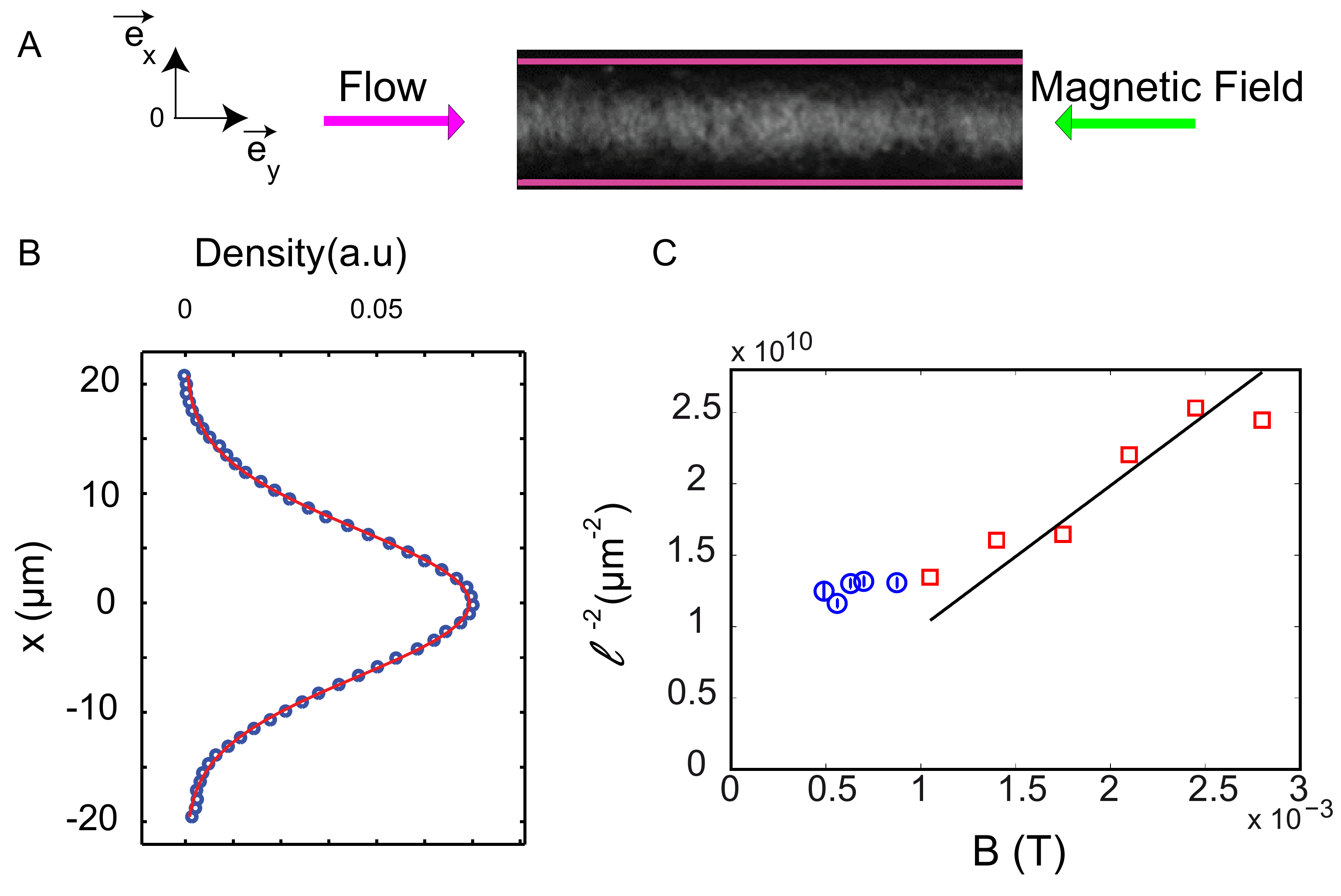}
    
    \caption{Flow focused state when bacteria are driven upstream, with the swimming direction imposed by the magnetic field antiparallel to the fluid flow. A: 
Picture of the system in a stationary state when bacteria are flushed with a central fluid velocity \textbf{$V_\mathrm{flow}=90 \pm 20\mu {\rm m/s}$} in a 50 microns wide channel, and a magnetic field  ($B=2.8\,$mT) tends to orient them in the opposite direction. B: Corresponding transverse intensity profile, adjusted with a gaussian. C: Evolution of the width of the gaussian profile with the magnetic field, for $V_\mathrm{flow} = 90 \pm 20\mu {\rm m/s}$, the red squares are adjusted by a linear fit $\ell^{-2} = 1.0 \times 10^{13}  {\rm m^{-2} T^{-1}}\times B$. The blue circles correspond to the regime where the stationary orientation assumption is not anymore valid.  \label{fig:fig1} } 
\end{figure}

We are now  in a position to quantitatively investigate the interplay between magnetically-driven active matter and an external flow field. This is done using microfluidic systems consisting of a main $50\,\mu$m wide channel (see Supp. Mat.), where the response of the magnetotactic microswimmers to steady Poiseuille flow is explored. We focus in the following on the situation where we orient the magnetic microswimmers against the fluid flow, and along the channel axis. While the flowing suspension remains fully homogeneously distributed over the channel width in the absence of imposed magnetic field, we observe that it rapidly concentrates in a narrower jet at the center of the channel upon turning the field on (fig.~\ref{fig:fig1}A).
Note that bacteria accumulate at the channel walls when swimmers are magnetically oriented in the flow direction (not shown).

This behavior is reminiscent of similar focusing phenomena reported for gyrotactic or phototactic motile algal cells \cite{Kessler85, Garcia13}. However, so far no quantitative experimental testing of gyrotaxis could be carried despite its long lasting description \cite{Pedley2015}. Beyond the demonstration of this phenomenon for magnetotactic bacteria, a key aspect of our system is that the characterization of the focusing mechanism can be made fully quantitative and can be remotely and continuously controlled by the external field.  Starting from an homogeneous flowing suspension, the latter focuses within $0.2\,$s (of order $w/V_\mathrm{swim}$) in a stationary magnetotactic bacteria jet. By varying the magnitude of $\mathbf{B}$, it is possible to change the efficiency of the trapping: the width of the focused suspension decreases for increasing magnetic driving.

Neglecting the out-of-plane component of the swimmers' velocity, we consider a bidimensional problem characterized by a Poiseuille flow with a velocity  field $\mathbf{u}$ and a maximum velocity $V_\mathrm{flow}$ in a channel of width $2w$. 
The hydrodynamic torque acting on a bacterium is:

\begin{equation}
\boldsymbol{\Gamma}_\mathbf{hyd} = \frac{\left[(1/2)\boldsymbol{\nabla}\times\mathbf{u}\right] - \boldsymbol{\Omega}}{\mu_R},
\end{equation}
with $ \boldsymbol{\Omega}$ the bacterium angular velocity. For small enough flow velocities, a stationary orientation of the bacterium ($\Omega=0$) is possible at every location in the channel:
\begin{equation}
\Gamma_\mathrm{hyd} =  -K\frac{x}{w},\mathrm{\ with\ }K = \frac{V_\mathrm{flow}}{\mu_Rw},
\end{equation}

with $x$ the distance from the channel center. 

As for the magnetic torque, it can be expressed as a function of the transverse speed $\dot{x}$ given that $\mathbf{V}_\mathrm{\mathbf{swim}}=\dot{x}\,\mathbf{e_x}+\mathbf{V}_\mathrm{\mathbf{axial}}$, where $\mathbf{V}_\mathrm{\mathbf{axial}}$ is the axial speed:
 \begin{equation}
 \boldsymbol{\Gamma}_\mathbf{mag} = \mathbf{M}\times\mathbf{B} = -MB\frac{\dot{x}}{V_\mathrm{swim}},
 \end{equation}

In this counter-flow orientation for $\mathbf{B}$,  the combined action of the magnetic field and the hydrodynamical shear drives the bacteria toward the center of the channel:

\begin{equation}
0 = MB\frac{\dot{x}}{V_\mathrm{swim}} + K\frac{x}{w}.
\end{equation}

Generalising the torque balance to account for the rotational noise $\zeta$ which is responsible for the finite extension of the focused jet, the bacterium radial trajectories obey the over-damped Langevin equation:
\begin{equation}
0 = MB\frac{\dot{x}}{V_\mathrm{swim}} + K\frac{x}{w} + \zeta.
\end{equation}
Note that the low velocity conditions for stationary orientation amounts to $K<MB$: magnetic torque exceeds the maximum hydrodynamic torque arising at the channel walls.

This Ornstein-Uhlenbeck process is formally equivalent to the one of a damped brownian particle in a harmonic potential, the hydrodynamical torque $K$ playing here the role of a harmonic potential. Solving the resulting Fokker-Planck equation leads to a Gaussian transverse distribution of the suspension for the steady-state bacterium density (see Supplementary Materials):
\begin{equation}
d(x) \propto e^{-\frac{x^2}{2\ell^2}}
\label{densexp}
\end{equation} 
with a width $\ell$ defined by:
\begin{equation}
\ell= w\sqrt{\frac{V_\mathrm{swim}\   B_c}{V_\mathrm{flow} \ B}}.
\label{eq:eq1}
\end{equation}
%
%
%%%%%%%%%%%%%%%%%%%%%%%
%
%
%%%%%%%%%%%%%%%%%%%%%%%
%

In line with the experimental observation, the model predicts that the higher the magnetic field is, the stronger the focusing will be. The same is expected when increasing the velocity, as it orients the swimmers farther apart from the axial field direction, towards the channel center. More quantitatively, the predicted gaussian distribution is very well obeyed by the experimental bacterium density profiles as shown in figure \ref{fig:fig1}.B. Investigating the profile width $\ell$ as a function of  $B$, we moreover evidence the $1/\sqrt{B}$ evolution (fig.\ref{fig:fig1}.C). Note that focusing widths at low magnetic fields seem to level off and to depart from the high field behavior. This departure is expected to occur due to a breakdown of the stationary orientation assumption throughout the whole channel.  This requires $B>1.2\pm0.3\,$mT  (Supp. Mat.), in very good agreement with the observed departure (blue circles on fig.\ref{fig:fig1}C).

\begin{figure}
\centering
    \includegraphics[width=0.5\textwidth]{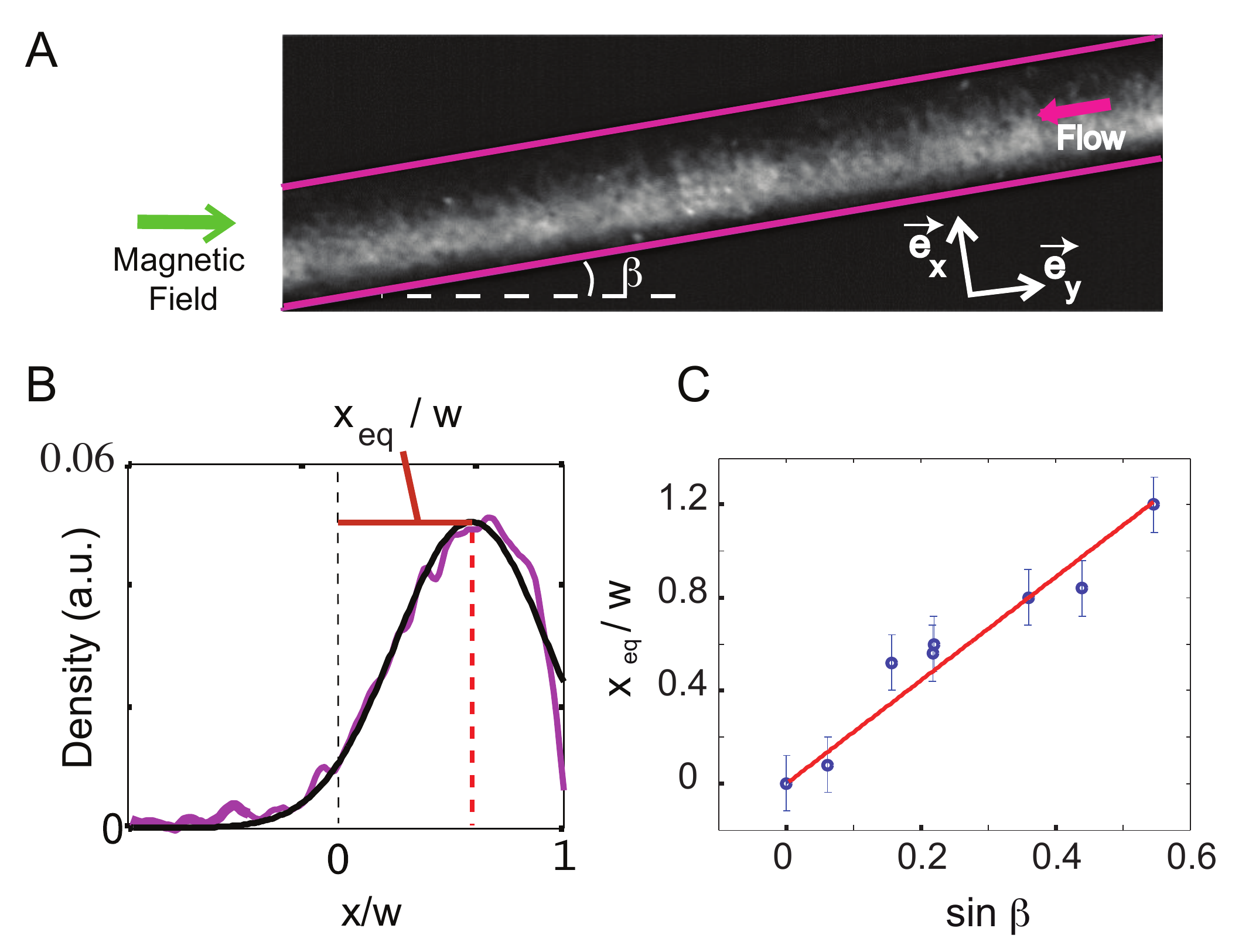}
    
    \caption{Moving the trap position. A: Sketch of the experiment, for $\beta = 8^\circ$, $w=50 \mu {\rm m}$ and $V_\mathrm{flow} = 50\mu {\rm m/s}$. B: Experimental density profile adjusted with a shifted gaussian. C: Dependency of the trap position $x_\mathrm{eq}/w$ with the tilting angle, for experiments with a magnetic field of 1.4mT and $V_\mathrm{flow} = 50 \mu {\rm m/s}$. We adjust the position with $x_\mathrm{eq}/w = A.\sin \beta$ and obtain $A = 2.1$. \label{fig:fig2}}. 
\end{figure}

In addition to the previous focusing process, it is also possible to adjust the jet location within the channel by tilting the magnetic field direction by an angle $\beta$ with respect to the channel axis. This is shown in fig.\ref{fig:fig2} A, where a suspension faces a central fluid velocity of half the typical swimming speed ($V_\mathrm{flow} = 50 \mu {\rm m/s}$), with a misalignment of an angle $\beta=8^\circ$. As can be seen, while still focused, the suspension jet is now shifted off-center, towards the channel side. Overall, the gaussian shape of the profile is maintained (fig.\ref{fig:fig2} B) with a peak offset $x_\mathrm{eq}$. Expanding on the previous theoretical description, the focusing location that corresponds to the expected maximum density reads (see Sup. Mat.):
\begin{equation}
 x_\mathrm{eq} = w\frac{MB}{K}\times\sin{\beta}.
 \end{equation}

Figure \ref{fig:fig2} C presents the evolution of the focusing position as a function of the misalignment $\beta$. The predicted linear dependency of $x_\mathrm{eq}$ in $\sin \beta$ is very well evidenced. This allows an experimental determination of $ MB/K = 2.1 \pm 0.2$.  The consistency of this value can be checked by estimating the associated hydrodynamic radius of the bacteria:
\begin{equation}
r=\left(\frac{Kw}{8\pi \eta V_\mathrm{flow}}\right)^{1/3}.
 \end{equation}

Considering a magnetic momentum $M=1 \pm 0.2 \times 10^{-16}\times {\rm A.m}^2$  \cite{Nadkarni2013,Wu2014},  $V_\mathrm{flow} = 50 \pm10 \mu {\rm m/s}$, $B=1.4\,$mT,  and $w=25 \pm 5 \,\mu$m, we obtain a hydrodynamic radius for bacteria: $r=1.1 \pm 0.2 \,\mu$m, which perfectly agrees with the size of the core bacteria, obtained with scanning electronic microscopy, around $1\,\mu$m.

%
%%%%%%%%%%%%%%%%%%%%%%%
%
%
%%%%%%%%%%%%%%%%%%%%%%%
%

Overall, the magnetotactic bacteria with their magnetic driving of physical nature, matching the Langevin paramagnetism, provide a model system of driven active microsystems, whose focusing interaction with non uniform flow can be fully quantitatively captured. Remarkably, this complex bio-systems is fully captured by a simple Fokker-Planck description. To our knowledge, this constitutes the first direct validation of such approaches initially proposed in the context of gyrotactic systems \cite{Pedley2015}, and are used as a key element for the description of complex bio-convection patterns and dynamics in such systems.

Note that such externally driven systems serve as conceptual building blocks for remotely controllable micro-surgeons or cargos that would be piloted through the blood vessel network. In that perspective, shifting the focused microswimmers beam towards the channel side where the flow slows down could be a strategy for achieving upstream swimming against high-speed blood flow. 

\subsection*{III. JET PEARLING TRANSITION}

\begin{figure}
\centering
    \includegraphics[width=0.4\textwidth]{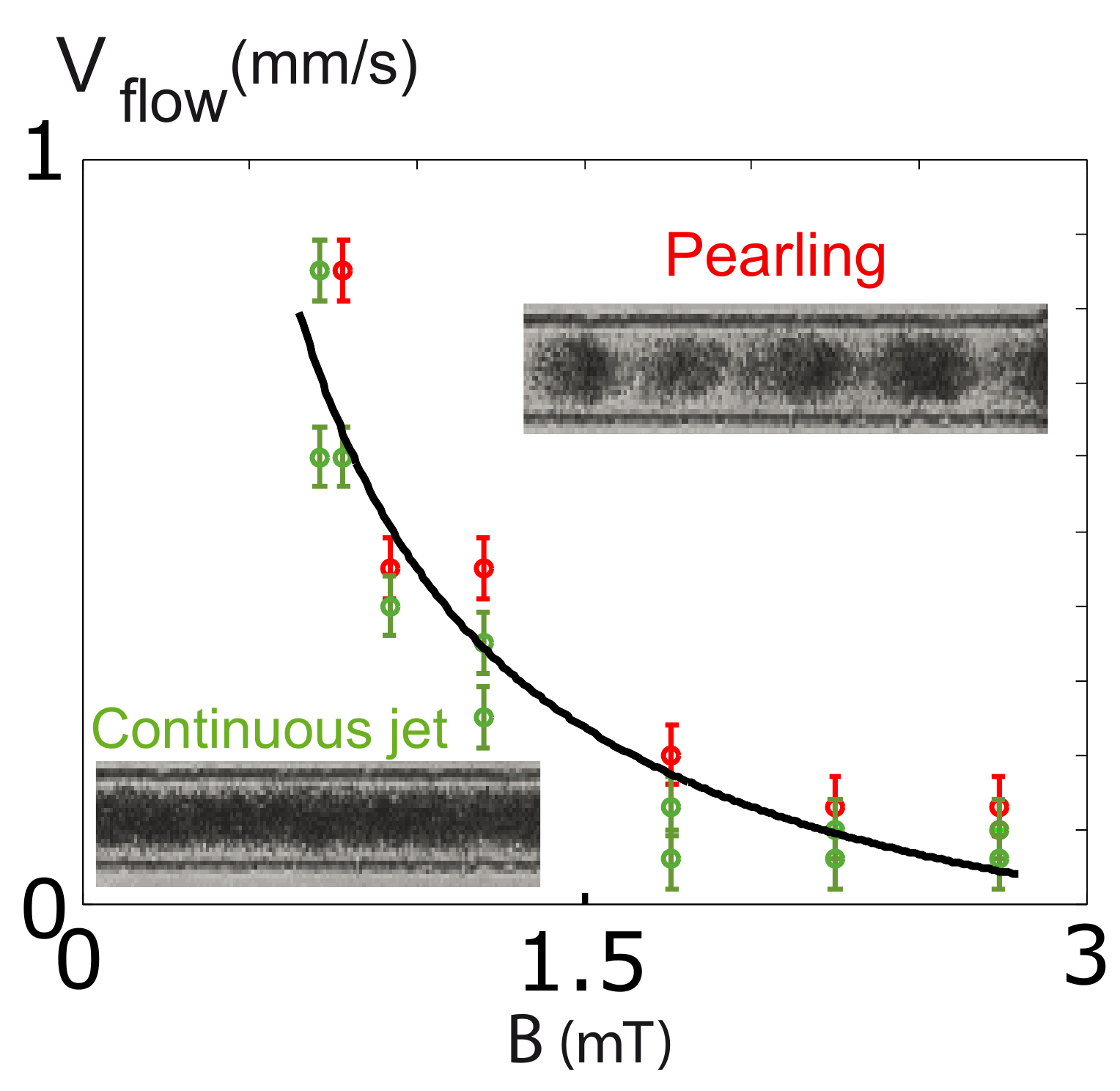}
    
    \caption{Phase diagram of the instability for various values of $V_\mathrm{flow}$ and $B$. Red dots are the unstable jets, green dots are the stable jets. The black line is the best adjustment with $V_\mathrm{flow}=a/B+c$ with $a=0.630$~mm.mT/s and $c=~-0.19$mm/s. The density in bacteria in this experiment is of the order of $10^7$ bacteria/$\mu$L. \label{fig:fig3}}
    
\end{figure}

So far, we have explored the interaction of driven magnetotactic bacteria and surrounding Poiseuille flows at moderate flow velocities. In this regime, a stationary suspension focusing was obtained, and increasing the magnetic field led to the concentration of the bacteria into a thinner jet. However, increasing further the external field, or increasing the flow velocity, induces a striking change of behavior. The originally stable beam of bacteria is destabilized into a pearling jet, which can yield to swarming droplets as seen in fig.\ref{fig:fig3}  (movies available in Supp. Mat.). Note that if we keep the same fluid flow but now lower the magnetic field, the focused suspension remains stable.

To carefully explore the $(V_\mathrm{flow},B)$ phase space, we fill the channel with magnetotactic bacteria, and for different imposed $B$, we vary the fluid flow $V_\mathrm{flow}$ oriented against the magnetic field. We then determine the conditions for which the focused jet remains stable or parts into suspension droplets.  This yields a phase diagram presented in fig.\ref{fig:fig3} showing two different dynamical regimes: a slow-flow/low-field regime corresponding to the previously described stable focused jet, and a fast-flow/high-field regime for which the suspension parts into droplets.

%
%%%%%%%%%%%%%%%%%%%%%%%%%%%
% 
%
%%%%%%%%%%%%%%%%%%%%%%%%%%%
% 
 
A natural argument for active systems --intimately associated with the onset of collective motions--, would be to relate the instability to the appearance of bacteria-bacteria interactions. Indeed, increasing the flow velocity or the magnetic field confines the bacteria in an increasingly smaller jet. Thus, the jet density goes like $\rho_0 (w/\ell)^2$, with $\rho_0$ the initially homogeneous bacteria density, and $\ell$ the jet radius as given by eq.~(\ref{eq:eq1}). Associating the transition to a threshold density one would expect a simple relationship between $B$ and $V_\mathrm{flow}$: $V^\mathrm{th}_\mathrm{flow}\propto {1}/{B_\mathrm{th}}$, in fair agreement with the experimental threshold (figure~\ref{fig:fig3}). Considering $\rho_0 \sim 10^7\,$bacteria/$\mu$L, this would translate into   a volume fraction of $\phi\sim32\,$\% in the jet. This suggests that 
 bacteria-bacteria interactions might play a role in the present pearling instability.

\begin{figure}
\centering
    \includegraphics[width=0.5\textwidth]{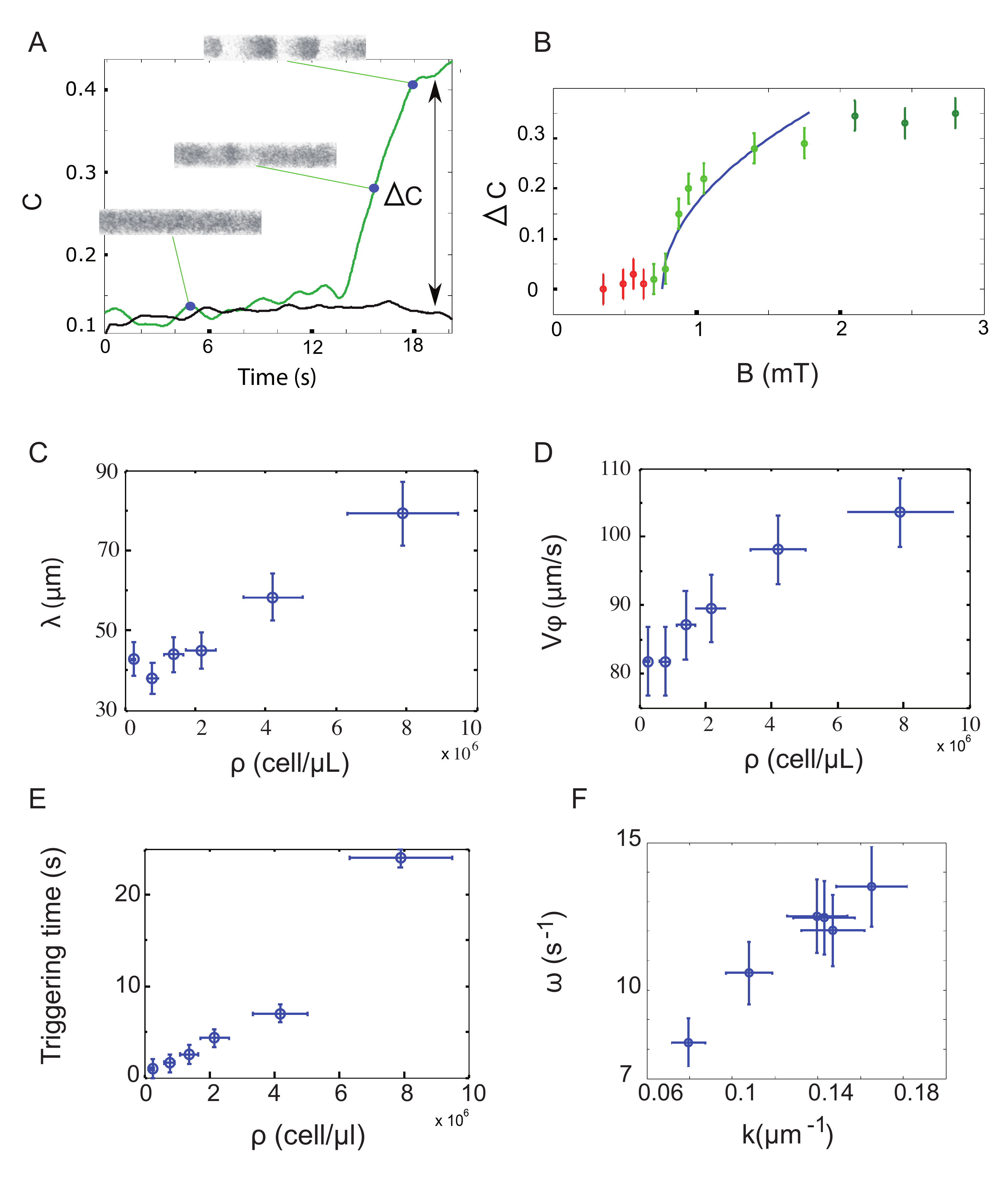}
    
    \caption{Characterisations of the instability . A: Evolution of the spatial contrast of the jet during flushing at $V_\mathrm{flow}=300\pm 40\,\mu$m/s, in green at 2.1 mT, in black at 0.3 mT. The blue dots correspond to the values of the parameters for which the pictures are taken.
    B: Evolution of the contrast burst $\Delta C$ with B for a central fluid velocity of $V_\mathrm{flow} = 300 \pm 40\,\mu$m/s, and the adjustment  by  $ (B-B_\mathrm{th})^{1/2}$
with $B_\mathrm{th} = 0.83\,$mT.
For figures C to F:  $B=2.8\,$mT and $V_\mathrm{flow}=180\,\mu$m/s. C: Evolution of the wavelength with the initial density. A net increase of the wavelength appears as the density increases. D. Evolution of the phase speed of the traveling wave in the laboratory frame as a function of the initial density.
    E: Evolution of the triggering time of the instability with the initial density.   F: Dispersion relation  $\omega$ as function of the wavenumber $k$.\label{fig:fig4}}
\end{figure}

We take benefit of the unprecedented control that offers the magnetotactic bacteria to provide some more experimental insights into the pearling transition characteristics. First, we  quantified the breakdown of the jet homogeneity
by increasing the magnetic field at a fixed velocity, set here to $V_\mathrm{flow} =300\pm40\,\mu$m/s. To characterize the modulation of the jet properties, we computed the mean axial density profile (over the field of view, typically $200\,\mu$m) $\hat{\rho}(y)=\langle \rho \rangle_x$ and defined the contrast parameter as:
\begin{equation}
C = \frac{\hat{\rho}_\mathrm{max}-\hat{\rho}_\mathrm{min}}{\hat{\rho}_\mathrm{max}+\hat{\rho}_\mathrm{min}},
\end{equation}
where $\hat{\rho}_\mathrm{max}$ (respectively  $\hat{\rho}_\mathrm{min}$) is the maximum (resp. the minimum) density along the channel.  Starting from a low magnetic field, stable jet situation, the target field is imposed at time $t$=0 and we look at the subsequent evolution of the jet contrast over time, as shown in fig.\ref{fig:fig4}~A.  For $B=0.35$ mT the contrast has a unique steady value over the duration of observation (black curve): there is an homogeneous and stationary focused beam of bacteria. The small drift observed is due to the small evolution with time of the density in bacteria. At a higher magnetic field of $B=2.1\,$mT, for which we are above the pearling instability threshold, the contrast shows a very distinct time evolution. Starting from a steady value corresponding to the focused beam, the contrast gradually increases until it plateaus again at a higher contrast ($+\Delta C$) corresponding here to the flow of the bacteria droplets. 
Figure \ref{fig:fig4} B presents the measured contrast increase as a function of the applied field $B$ with fixed flow velocity. The transition from stable flow focused state to pearling jet is very clearly revealed at $B_\mathrm{th}= 0.83\,$mT, with $\Delta C$ departing from zero and further increasing as $B$ gradually moves above the threshold. Qualitatively, this looks like a supercritical Hopf bifurcation between a stable homogeneous jet and a varicose modulation. This suggests that the modulation amplitude might evolve like the square root of the distance to threshold, in fair agreement with the experimental behavior as shown by the adjustment of data by
\begin{equation}
 \Delta C \propto (B-B_\mathrm{th})^{1/2}.
\end{equation}
%
%%%%%%%%%%%%%%%%%%%%%%%%%%
%

To carry on with experimental characterizations of the pearling instability, we performed additional measurements as function of the bacterial density.
We have used a bacteria train of slowly varying density (see Suppl. Mat. for experimental details), from which two different information have been extracted.   

First, while the focusing time for creating the bacteria jet is always quick ($< 0.3\,$s),  the triggering time for the suspension to destabilize can be as large as $>20\,$s. More quantitatively, we observe (fig.\ref{fig:fig4} E) that the triggering time increases with bacteria density for fixed values of magnetic field and Poiseuille flow (here:  $B=2.8\,$mT and  $V_\mathrm{flow}=180\,\mu$m/s). Note that on the contrary, we did not observe a change of triggering time for fixed density but variable magnetic field amplitudes. Second, we have also measured the characteristic length scale of the instability pattern as a function of the density. As shown in fig.~\ref{fig:fig4}~C this wavelength increases with the initial density.

Measuring the phase speed of the jet modulation (fig.\ref{fig:fig4} D), we are also able to obtain a portion of the dispersion relation for the instability (fig.\ref{fig:fig4} F). 
The phase speed appears quite different from the Poiseuille central velocity (here $V_\mathrm{flow}$ =$180\,\mu$m/s) and much more compatible with an almost fixed modulation in a frame travelling at $V_\mathrm{flow}-V_\mathrm{swim}$.

\begin{figure}
\centering
    \includegraphics[width=0.25\textwidth]{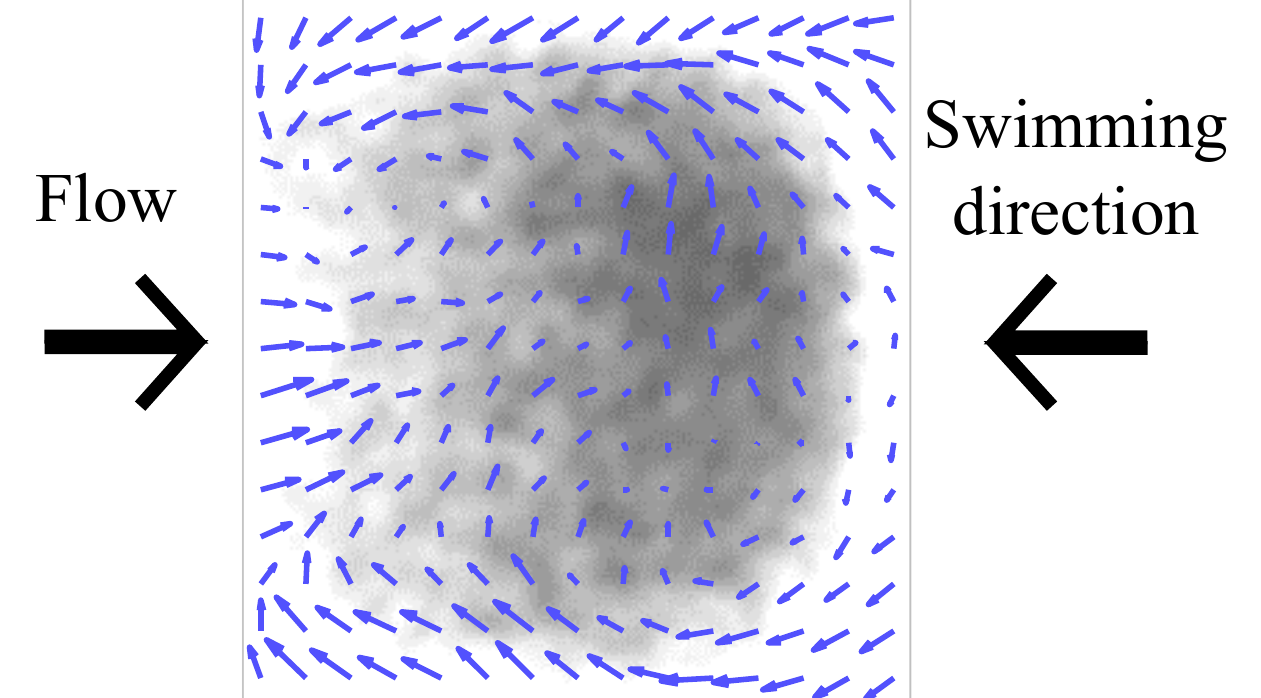}
    
    \caption{A tracked swarming droplet, with the superposed bacteria velocities, one can easily distinguish a recirculation inside the droplet.\label{fig:fig6}}
    
\end{figure}

As a final experimental characterization of the instability, we report in figure~\ref{fig:fig6} the relative velocities of bacteria into a swarm obtained once the suspension has fully segregated. 
We observe an inner circulation composed of two contra-rotative bacteria vortices, showing the complexity of the coupling between the flow and the bacteria motion in good agreement with the dependency of the phase speed with the density exhibited in figure~\ref{fig:fig4}.D. This suggests a feedback of the active suspension on the fluid velocity.

At very high density, one can expect that steric interactions will refrain this recirculation to happen, and hence slow down the destabilization process.
Quite remarkably  the density of the interacting objects increases the size of the final pattern obtained, but  the dynamics to obtain this pattern is slowed down by the density.\

%
%%%%%%%%%%%%%%%%%%%%%
%
% 
 
 \subsection*{IV. DISCUSSION: CHALLENGING THEORIES}
 
A theoretical description of this instability requires identifying the relevant ingredients involved in this complex system: a driven active system under flow. Of course liquid jet instabilities are known to occur also in purely \textit{passive} (i.e. equilibrium) systems: we can think for instance to Kelvin-Helmoltz or Rayleigh-Plateau \cite{Eggers2008}. However, thinking along these lines would require defining for the effective active fluid physical properties such as viscosity or surface tensions. While viscosity of such systems have been studied \cite{Saintillan2009, Sokolov2009, Rafai2010}, the idea of generalizing the surface tension in active systems has only emerged very recently \cite{Bialke2015}. Indeed the mere notion of active pressure was recently introduced \cite{Takatori2014,Yang2014,Solon2015} and here with field oriented swimmers active pressure anisotropy and active surface tensions might be expected. However, the practical meaning and usefulness of these different variables is still a matter of extensive theoretical research \cite{Solon2015b} that precludes its current use for drawing generalization from passive systems instabilities. 

Nevertheless, active matter primarily attracted attention because of the spontaneous appearance of collective motions associated to spatio-temporal coherent structures, for instance in various biological micro-organisms \cite{Ishikawa2009}. Thus instabilities and pattern formation specific to active systems have been the subject of intense researches lately \cite{Hill2005,Koch2011,Ezhilan2013}. Of particular interest here are gravitactic and phototactic systems which share with magnetotaxis the ability of being remotely oriented by an external field. 

Recent experiments on phototactic 
swimmers \cite{Garcia13} indeed reported the focusing regime, yet without noticing jet instabilities. A following numerical paper, considering puller phototactic swimmers submitted to a Poiseuille flow showed a pearling instability \cite{Jibuti15}. However, despite this reported instability was assigned to hydrodynamic interactions between swimmers, it seems to depart from the present one. Indeed no instability threshold ---meaning no stable focused-jet--- is found numerically. Additionally, the self-focusing time is found to be comparable with the subsequent time for jet breaking into swimmers' pearls while these timescales are quite apart in the present study. Moreover, besides the phenomenology it only gives a qualitative picture for the instability based on the puller-puller interactions.

Concerning gravitactic bio-organisms, they have long been known to exhibit gyrotaxis \cite{Kessler85} where gravity plays a similar role to the present magnetic field and where bottom-heavy cells orients by a balance between gravitational and shear torques. There, self-focused jets have been observed that might give rise to blip instability as reported experimentally \cite{Kessler85,Denissenko2007}. Here again, the instabilities do not seem to have the same behaviors and properties. While a velocity threshold is now also observed, gyrotactic swimmers display a restabilization of the jet at high flow velocities, in strong contrast to our findings. 
The phase speed of the jet modulation in gyrotactic swimmers 
was also reported to be close to the central velocity \cite{Kessler85,Denissenko2007}, which is not what we observe with our system.

Qualitatively also: magnetotactic instability looks like a varicose mode with definite characteristic length scales evolving toward isolated swarms, while gyrotactic algae were shown to yield very distant concentrated drops of swimmers connected with thin filament, or radius transition between coexisting jets \cite{Denissenko2007}. 

On the theoretical side much has been done on gyrotactic instabilities \cite{Hill2005,Koch2011,Ezhilan2013} owing to its relevance for bio-convection, organic matter plumes, etc. \cite{Kessler85,Platt61,Kessler86}. However, the presently reported instability for magnetotactic bacteria lacks a key ingredient usually involved in the previously mentioned phenomenology. Namely, gyrotactic systems do incorporate a gravitational body force which couples the flow field with the local micro-organisms density, due to the density mismatch. For magnetotactic bacteria, it is important to stress that no such equivalent exist as \textit{no force is exerted on the bacteria by the field}. In recent studies of jet stability of gyrotactic swimmers  \cite{Hwang2014}, this body-force term is found to have a major importance for the explored instabilities. 
 Therefore, direct comparison with our system is  difficult.
  Moreover, discussion is made in the absence of swimmers interaction --the swimming stress contribution \cite{Simha2002, Saintillan2008, Hwang2014} have been neglected-- while a simple analysis of the instability thresholds suggests here that it plays a significant role.

To summarize, we have unveiled new rich behaviors of driven active matter submitted to an external field. We have studied a novel model of active system, made of magnetotactic bacteria, whose motion can be tuned and oriented by an external magnetic field.  When facing a Poiseuille flow, at moderate velocity and low magnetic field,  the suspension focuses in the center of the channel, what reminisces about similar behaviors observed for phototactic or gravitactic systems. We propose a simple stochastic model which allows to quantitatively predict the strength and the position of the trap as a function of the direction and the magnitude of the external driving field. This is, to our knowledge, the first time that a quantitative description of the trapping efficiency of driven active matter has been achieved. Providing a quantitative description of those systems is a key element for building up models  to understand the complex patterning behaviors of microswimmers colonies.  For higher values of the magnetic field or the velocity of the imposed flow we observe a transition towards a new state where the bacterial suspension destabilizes in swarming droplets. This instability illustrates new structuring capabilities of driven active matter and challenges actual theoretical frameworks of active matter, suggesting further horizons to explore in this extremely rich domain.
 
\begin{acknowledgments}
We thank Siddardha Koneti for his work on the instability, Veroniquie Utzinger, Guy Condemine and Nicole Cotte-Pattat for their help with the bacterial cultures. We thank
INL for access to their clean room facilities and AXA research fund for its financial support. 
\end{acknowledgments}

%
%%%%%%%%%%%%%%%%%%%%%%%%%%%%%%%%%%%%%
%
%

\end{document}